\documentclass[journal,twocolumn]{IEEEtran}
\usepackage{amssymb}
\usepackage{amsfonts}
\usepackage{amsmath}
\usepackage{algorithm}
\usepackage{algorithmic}

\usepackage{graphicx,subfigure,amsmath,amssymb,cite}

\usepackage[dvips]{color}
\usepackage{float}
\ifCLASSINFOpdf
\else
\fi
\begin{document}
\title{GPI-based Secrecy Rate Maximization Beamforming Scheme for Wireless Transmission with AN-aided Directional Modulation}

\author{Hai Yu, Simin Wan, Wenlong Cai, Ling Xu,   Xiaobo Zhou, Jin Wang, Yongpeng Wu, \\Feng Shu, Jiangzhou Wang, and Jianxin Wang

% <-this % stops a space
\thanks{This work was supported in part by the National Natural Science Foundation of China (Grant Nos. 61771244, 61702258, 61472190, 61501238 and 61602245), the China Postdoctoral Science Foundation (2016M591852), the Postdoctoral research funding program of Jiangsu Province(1601257C), the Natural Science Foundation of Jiangsu Province (Grants No. BK20150791), and the open research fund of National Mobile Communications Research Laboratory, Southeast University, China (No.2013D02).}% <-this % stops a space

\thanks{Hai Yu, Simin Wang, Ling Xu, Feng Shu,  Xiaobo Zhou, Jin Wang,  and Jianxin Wang are with School of Electronic and Optical Engineering, Nanjing University of Science and Technology, Nanjing, 210094, China.}
\thanks{Wenlong Cai is with the faculty of National Key Laboratory of Science and Technology on Aerospace Intelligence Control and Beijing Aerospace Automatic Control Institute, Beijing, China. E-mail: caiwenlon@buaa.edu.cn}
\thanks{Yongpeng Wu is with the Shanghai Key Laboratory of Navigation and Location-Based Services, Shanghai Jiao Tong University, Minhang 200240, China. E-mail: yongpeng.wu2016@gmail.com}
\thanks{Feng Shu is also with the College of Computer and Information Sciences, Fujian Agriculture and Forestry University, Fuzhou 350002, China.}
\thanks{Jiangzhou Wang is with the School of Engineering and Digital Arts, University of Kent, Canterbury CT2 7NT, U.K. E-mail: j.z.wang@kent.ac.uk}
}

%\markboth{Journal of \LaTeX\ Class Files,~Vol.~6, No.~1, January~2007}%
%{Shell \MakeLowercase{\textit{et al.}}: Bare Demo of IEEEtran.cls for Journals}



\maketitle

\begin{abstract}
In a directional modulation network, a general power iterative (GPI) based beamforming scheme is proposed to maximize the secrecy rate (SR), where there are two optimization variables required to be optimized. The first one is the useful precoding vector of transmitting confidential messages to the desired user while the second one is the artificial noise (AN) projection matrix of forcing more AN to eavesdroppers. In such a secure network, the paramount problem is how to design or optimize the two optimization variables by different criteria. To maximize the SR (Max-SR), an alternatively iterative structure (AIS) is established between the AN projection matrix and the precoding vector for confidential messages.  To choose a good initial value of iteration process of GPI, the proposed Max-SR method  can readily double its convergence speed compared to the random choice of initial value. With only four iterations, it may rapidly converge to its rate ceil. From simulation results, it follows that the SR performance of the proposed  AIS of GPI-based Max-SR is much better than those of conventional leakage-based and null-space projection methods in the medium and large signal-to-noise ratio (SNR) regions, and its achievable SR performance gain  gradually increases as SNR increases.

\end{abstract}
% IEEEtran.cls defaults to using nonbold math in the Abstract.e


%\newpage

 %\ifCLASSOPTIONpeerreview
% \fi

\IEEEpeerreviewmaketitle

%\section{INTRODUCTION}
%\newpage

\section{Introduction}
In the recent decade,  physical-layer security in wireless networks, as a new tool to provide an incremental safeguard of confidential message over conventional cryptography, has drawn tremendous research attention and interests from both academia and industry \cite{Wyner1975,zhao,YAN1,zou1,Li1,hmwang,Lv,Hanif,Han1,Han2,Han3}. In \cite{Wyner1975}, the author's seminal research work has established the  channel model and found the tradeoff curve between the transmission rate  and the data equivocation  seen by the wire-tapper. More importantly, the author also proved that reliable transmission at rates up to C, is possible in approximately perfect secrecy. As a physical layer secure transmit technique suitable for line-of-propagation (LoP) scenario, directional modulation (DM) has made great progresses in many aspects  with the aid of artificial noise (AN) and antenna array beamforming\cite{maha,kalantari,babakhani,daly,Ding,Hu1,Wu,Zhu}. To enhance security,  the symbol-level precoder in \cite{kalantari} was presented by using the concept of constructive interference in directional modulation with the goal of reducing the energy consumption at transmitter. In the presence of  direction measurement error,  the authors in \cite{Hu1,Wu} designed three  new robust DM synthesis methods for three different application scenarios: single-desired user, multi-desired user broadcasting, multi-desired user MIMO  by fully making use of the property of direction measurement error. Compared with conventional non-robust methods, like null-space projection (NSP) \cite{Ding,Dingy}, the proposed robust methods actually harvests an appealing rate gain and  almost an order-of-magnitude bit error rate (BER) performance improvement along desired directions. {\color{blue}{The works above only present an investigation on conventional DM with only direction-dependent property. However, if eavesdropper lies on the same direction from as the desired user, and its distance from the DM transmitter is different from the distance from the desired user to the DM transmitter, then it can still intercept the confidential messages successfully. This is the existing intrinsic secure problem for the conventional DM networks. To address the serious secure issue, recently, the authors in \cite{Hu, ShuF}  proposed an original concept of secure and precise wireless transmission. In their works, the confidential messages are transmitted to the desired position precisely and securely.  If eavesdropper is outside the small area around the desired position and locates in the same direction as the desired direction, it can not intercept the confidential messages successfully due to  AN corruption and frequency random property. However, in this paper,  we focus still on how  to maximize the SR for conventional DM networks via GPI algorithm.}}

%In general, due to its directional property, the DM in LoP channel might harvest a high array gain along the desired directions via confidential-message beamforming and  significantly worsen the performance of eavesdroppers at undesired directions by forcing AN to the tap-wire direction. Consequently,  the resulting goal of secure transmission is reached successfully.

 However, in a three-node directional modulation network, what is the maximum achievable secrecy rate in the absence of direction estimation error? This is an open NP-hard  problem. In the following, we will solve the problem via a combination of general power iterative (GPI) method \cite{Lee} and alternatively iterative structure (AIS). From simulation results, we find that the proposed GPI-based AIS can achieve an obvious secrecy rate (SR) performance gain compared to conventional NSP and leakage-based methods in the medium and large SNR regions. {\color{blue}{Our main contributions are summarized as follows:
 \begin{enumerate}
  \item
 In line-of-sight (LoS) scenarios, such as mmWave communications (massive MIMO), internet of things (IoT), unmanned aerial vehicle (UAV) and satellite communications,  we propose a general power iterative (GPI) scheme in DM system to maximize secrecy rate (SR), which is shown to be much better than conventional NSP and leakage-based methods in terms of SR.
  \item To accelerate the convergence rate of our proposed method,  the leakage-based precoding vector and AN projection matrix are chosen to be the initial values instead of random generating. From simulation results, it follows that the new initial values can greatly reduce the number of iterations, and doubles the convergence rate compared to the random ones. This means that a good choice of initial values leads to a dramatic reduction  in computational amount of the  proposed method.
\end{enumerate}}}
%Authors in \cite{maha, kari1} established a unified framework for physical layer multi-casting to multiple co-channel groups, where multiple independent data streams are transmitted to groups of users by the multiple antennas. If one group of eavesdropper appears, then how to achieve a secure transmission in such a situation is an interesting and important research topic. This secure problem includes twofold:  the privacy protection among desired groups and the leakage of all confidential messages from all desired groups to the group of eavesdroppers. We will address this topic from the standpoint of physical layer security by using DM  in this paper, where the LoP channel is considered and  two precoding and AN projecting schemes for this multi-cast communications will be proposed.

The remainder is organized as follows: Section II describes the system model. An AIS of maximizing secrecy rate (Max-SR) based on GPI with aided AN is proposed in Section III.  Section IV presents the simulation results and complexity analysis. Finally, we draw conclusions in Section V.

Notations: throughout the paper, matrices, vectors, and scalars are denoted by letters of bold upper case, bold lower case, and lower case, respectively. Signs $(\cdot)^T$, $(\cdot)^*$£¬ and $(\cdot)^H$ denote transpose,~conjugate,~and conjugate transpose,~respectively. Notation $\mathbb{E}\{\cdot\}$ stands for the expectation operation. Matrices $\textbf{I}_N$ denotes the $N\times N$ identity matrix and $\textbf{0}_{M\times N}$ denotes $M\times N$ matrix of all zeros. $\text{tr}(\cdot)$ denotes matrix trace. Operation $\otimes$ denotes the Kronecker product of two matrices  \cite{Horn}.

\section{System Model}
\begin{figure}[h]
 \centering
 %Requires \usepackage{graphicx}
 \includegraphics[width=0.4\textwidth]{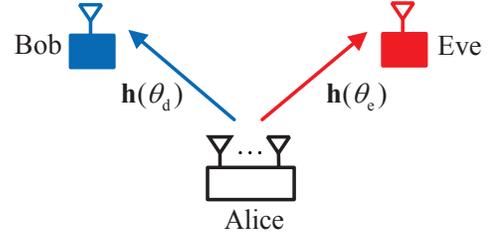}\\
 \caption{Directional modulation network}\label{Sys_Mod}
\end{figure}

 Fig.~\ref{Sys_Mod} sketches a diagram block of directional modulation (DM) network consisting of one base station (BS, Alice) equipped with $N$ antennas, one desired node (Bob) and one eavesdropping node (Eve). Here, it is assumed that both desired node and eavesdropping node are employed with single antenna and BS employs an $N$-element uniformly spaced linear array. Due to its directional property,  DM usually works in LoP channel. By introducing the precoding vector of confidential messages and AN projection matrix at transmitter,  the transmit baseband signal from antenna array of BS is of the form
\begin{equation}\label{Tx signal s}
\mathbf{s}=\beta_1\sqrt{P_s}\mathbf{v}_dx+\alpha\beta_2\sqrt{P_s}\mathbf{P}_{AN}\mathbf{z},
\end{equation}
where $x$ is the confidential message with $\mathbb{E}\left\{x^Hx\right\}=1$ and $\mathbf{z}\in\mathbb{C}^{N\times1}$ denotes the AN vector of obeying complex Gaussian distribution $\mathcal{C}\mathcal{N}(0,\mathbf{I}_{N-1})$. $P_s$ denotes the total transmit power, $\beta_1$ and $\beta_2$ stand for the power allocation (PA) factors of confidential message and AN with $\beta^2_1+\beta^2_2=1$. $\mathbf{v}_d\in\mathbb{C}^{N\times1}$ denotes the transmit beamforming vector to align the confidential message to desired direction, $\mathbf{v}_d^H\mathbf{v}_d=1$. $\mathbf{P}_{AN}\in\mathbb{C}^{N\times (N-1)}$ is the projection matrix, $\alpha$ normalizes $\mathbf{P}_{AN}\mathbf{z}$ such that $\alpha^2\mathbb{E}\{\mathrm{Tr}[\mathbf{P}_{AN}\mathbf{z}\mathbf{z}^H\mathbf{P}_{AN}^H]\}=1$.
%, which means $\mathrm{Tr}[\mathbf{P}_{AN}\mathbf{P}_{AN}^H]=\frac{1}{\alpha^2}$.

After experiencing the LoP channel, the received signal along direction $\theta$ is given by
\begin{align}\label{Rx_signal y}
y(\theta)
&=\mathbf{h}^{H}(\theta)\mathbf{s}+n_r\nonumber\\
&=\beta_1\sqrt{P_s}\mathbf{h}^{H}(\theta)\mathbf{v}_dx+\alpha\beta_2\sqrt{P_s}\mathbf{h}^{H}(\theta)\mathbf{P}_{AN}\mathbf{z}+n_r,
\end{align}
where $n_r$ is additive white Gaussian noise (AWGN) with $n_r\sim\mathcal{C}\mathcal{N}(0,\sigma_r^2)$, $\mathbf{h}(\theta)\in\mathbb{C}^{N\times1}$ is the normalized steering vector defined by
%\begin{align}\label{steer}
$\mathbf{h}(\theta)=\sqrt{N^{-1}}[\exp\left(j2\pi\Psi_{\theta}(1)\right),$ $ \cdots,  \exp\left(j2\pi\Psi_{\theta}(N)\right)]^T$,
%\end{align}
and the phase function $\Psi_{\theta}(n)$ along direction $\theta$ is denoted as
%\begin{equation}\label{var_phi}
$\Psi_{\theta}(n)\triangleq-[n-(N+1)/2]d\cos\theta\lambda^{-1}, (n=1,2,\cdots,N)$,
%\end{equation}
where $n$ indexes the elements of transmit antenna array, $d$ denotes the spacing of two adjacent antennas, and $\lambda$ is the wavelength of the carrier. The receive signals at Bob and Eve are written as
\begin{align}\label{Rx_signal yd}
y(\theta_d)
%&=\mathbf{h}^{H}(\theta_d)\mathbf{s}+n_d\nonumber\\
&=\beta_1\sqrt{P_s}\mathbf{h}^{H}(\theta_d)\mathbf{v}_dx+\alpha\beta_2\sqrt{P_s}\mathbf{h}^{H}(\theta_d)\mathbf{P}_{AN}\mathbf{z}+n_d,
\end{align}
and
\begin{align}\label{Rx_signal ye}
y(\theta_e)
%&=\mathbf{h}^{H}(\theta_e)\mathbf{s}+n_e\nonumber\\
&=\beta_1\sqrt{P_s}\mathbf{h}^{H}(\theta_e)\mathbf{v}_dx+\alpha\beta_2\sqrt{P_s}\mathbf{h}^{H}(\theta_e)\mathbf{P}_{AN}\mathbf{z}+n_e,
\end{align}
where $n_d\sim\mathcal{C}\mathcal{N}(0,\sigma_d^2)$ and $n_e\sim\mathcal{C}\mathcal{N}(0,\sigma_e^2)$ represent the noise at Bob and Eve. {\color{blue}{With the help of some classic estimation algorithms such as Capon's method, MUSIC and ESPRIT, the DM transmitter in Fig.~1 can obtain the direction knowledge of eavesdroppers and desired users.}} Without loss of generality, it is assumed that the variances of $\mathbf{\sigma}_r^2$ ,~$\mathbf{\sigma}_d^2$,~and $\mathbf{\sigma}_r^e$ are equal, i.e., $\sigma_r^2=\sigma_d^2=\sigma_e^2=\sigma^2$.

To derive the SR,  the achievable rates of Bob and Eve, $R(\theta_d)$ and $R(\theta_e)$ are defined as follows
\begin{align}\label{Rd}
R(\theta_d)
% & \triangleq I(y(\theta_d);[x,\theta_d])\nonumber\\
 & = \log_2\left(1+\frac{\beta_1^2P_s \mathbf{h}^H(\theta_d)\mathbf{v}_d\mathbf{v}_d^H\mathbf{h}(\theta_d)}{\sigma^2_d+\alpha^2\beta_2^2P_s \mathbf{h}^H(\theta_d)\mathbf{P}_{AN}\mathbf{P}_{AN}^H\mathbf{h}(\theta_d)}\right),
\end{align}
and
\begin{align}\label{Re}
R(\theta_e)
 %& \triangleq I(y(\theta_e);[x,\theta_e])\nonumber\\
 & = \log_2\left(1+\frac{\beta_1^2P_s \mathbf{h}^H(\theta_e)\mathbf{v}_d\mathbf{v}_d^H\mathbf{h}(\theta_e)}{\sigma^2_e+\alpha^2\beta_2^2P_s \mathbf{h}^H(\theta_e)\mathbf{P}_{AN}\mathbf{P}_{AN}^H\mathbf{h}(\theta_e)}\right),
\end{align}
%where $I(y(\theta);[x,\theta])$ means the mutual information along $\theta$ between the input symbol $x$ and the output symbol $y$.
Then, the SR $R_s(\mathbf{v}_d,\mathbf{P}_{AN})$ is defined as follows
{\color{blue}
\begin{align}\label{Rs1}
R_s(\mathbf{v}_d,\mathbf{P}_{AN})
 &=\max\{0,R(\theta_d)-R(\theta_e)\}\nonumber\\
 %&=\log_2\frac{1+\frac{\beta_1^2P_s \mathbf{h}^H(\theta_d)\mathbf{v}_d\mathbf{v}_d^H\mathbf{h}(\theta_d)}{\sigma^2_d+\alpha^2\beta_2^2P_s \mathbf{h}^H(\theta_d)\mathbf{P}_{AN}\mathbf{P}_{AN}^H\mathbf{h}(\theta_d)}}{1+\frac{\beta_1^2P_s \mathbf{h}^H(\theta_e)\mathbf{v}_d\mathbf{v}_d^H\mathbf{h}(\theta_e)}{\sigma^2_e+\alpha^2\beta_2^2P_s \mathbf{h}^H(\theta_e)\mathbf{P}_{AN}\mathbf{P}_{AN}^H\mathbf{h}(\theta_e)}}\nonumber\\
&=\max\{0,\log_2\left(\frac{\mathbf{v}_d^H(\mathbf{H}_d+A_d\mathbf{I}_N)\mathbf{v}_d}{\mathbf{v}_d^H(\mathbf{H}_e+A_e\mathbf{I}_N)\mathbf{v}_d}\times B\right)\},
 \end{align}}
where $\mathbf{H}_d\triangleq\mathbf{h}_d\mathbf{h}_d^H$, $\mathbf{H}_e\triangleq\mathbf{h}_e\mathbf{h}_e^H$,
\begin{align}
A_d=\alpha^2\beta_2^2\beta_1^{-2}\mathbf{h}^H(\theta_d)\mathbf{P}_{AN}\mathbf{P}_{AN}^H\mathbf{h}(\theta_d)+\sigma^2(\beta_1^2P_s)^{-1},
\end{align}
\begin{align}
A_e=\alpha^2\beta_2^2\beta_1^{-2}\mathbf{h}^H(\theta_e)\mathbf{P}_{AN}\mathbf{P}_{AN}^H\mathbf{h}(\theta_e)+\sigma^2(\beta_1^2P_s)^{-1},
\end{align}
and
\begin{align}
B=\frac{\mathbf{h}^H(\theta_e)\mathbf{P}_{AN}\mathbf{P}_{AN}^H\mathbf{h}(\theta_e)+\sigma^2(\alpha^2\beta_2^2P_s)^{-1}}  {\mathbf{h}^H(\theta_d)\mathbf{P}_{AN}\mathbf{P}_{AN}^H\mathbf{h}(\theta_d)+\sigma^2(\alpha^2\beta_2^2P_s)^{-1}}.
\end{align}
{\color{blue}{The RS defined in (7) should be larger than or equal to zero. If it is less than zero, then the eavesdropper will get more mutual information than the desired user. This is not desired in a secure system.   However, the second term on right hand side of (7) can be less than zero sometimes, i.e., \\
\begin{align}
\log_2\left(\frac{\mathbf{v}_d^H(\mathbf{H}_d+A_d\mathbf{I}_N)\mathbf{v}_d}{\mathbf{v}_d^H(\mathbf{H}_e+A_e\mathbf{I}_N)\mathbf{v}_d}\times B\right)<0
\end{align}
if the eavesdropper is closer to the DM transmitter than the desired user. Thus, the operation max in RS given by (7) should be kept such that the SR is larger than or equal to zero.}}
In what follows, we will maximize the SR by optimizing the beamforming vector $\mathbf{v}_d$ of confidential messages and the projection matrix $\mathbf{P}_{AN}$ of AN, which can be casted as
\begin{align}\label{P1}
\mathrm{(P1):}&\max_{\mathbf{v}_d,\mathbf{P}_{AN}}~~~~R_s(\mathbf{v}_d,\mathbf{P}_{AN})\\ \nonumber
&~~\text{subject to}~~\mathbf{v}_d^H\mathbf{v}_d=1, \mathrm{Tr}[\mathbf{P}_{AN}\mathbf{P}_{AN}^H]=\alpha^{-2}.
\end{align}
Obviously, the above optimization problem  is  NP-hard, and it is hard to be solved directly or find a closed-from solution to the above problem.{\color{blue}{The objective function in (\ref{P1}) as shown in (\ref{Rs1}) is divided into a sum of two non-convex functions of variables $\mathbf{v}_d$, and $\mathbf{P}_{AN}$. This means that it is still a non-convex function. Thus, we make a conclusion  that the optimization problem  in (\ref{P1}) is a nonlinear non-convex optimization.}}

\section{Proposed Iterative algorithm to design beamforming vector $\mathbf{v}_d$ and projection matrix $\mathbf{P}_{AN}$}
Since the joint optimization $\mathbf{v}_d$ and $\mathbf{P}_{AN}$ in (\ref{P1}) is too complicated, we divided it into two mutual coupling subproblems and construct an iterative structure between them. By fixing the beamforming vector $\mathbf{v}_d$, the optimal $\mathbf{P}_{AN}$ is solved by utilizing the GPI algorithm \cite{Lee}. For given projection matrix $\mathbf{P}_{AN}$, the optimal $\mathbf{v}_d$ is derived according to generalized Rayleigh-Ritz ratio in \cite{Horn}. This forms an alternatively iterative structure. The Max-SR method outputs  the beamforming vector $\mathbf{v}_d$ and projection matrix $\mathbf{P}_{AN}$ by repeatedly applying the GPI in \cite{Lee}.
\subsection{Optimize $\mathbf{P}_{AN}$ for fixed $\mathbf{v}_d$}
If the beamforming vector $\mathbf{v}_d$ in (\ref{P1}) is fixed, then the optimization problem (\ref{P1}) can be reduced to
\begin{align}\label{P1_1}
\mathrm{(P1.1):}&\max_{\mathbf{P}_{AN}}~~~~~~~R_s(\text{fixed}~\mathbf{v}_d,\mathbf{P}_{AN})\\ \nonumber
&\text{subject to}~~~\mathrm{Tr}[\mathbf{P}_{AN}\mathbf{P}_{AN}^H]=\alpha^{-2}.
\end{align}
The cost function  $R_s(\text{fixed}~\mathbf{v}_d,\mathbf{P}_{AN})$ is rewritten as

\begin{align}\label{Rs2}
R_s
&=\log_2\left(\frac{\mathrm{Tr}(\mathbf{P}_{AN}^H\mathbf{B}_d\mathbf{P}_{AN})}{\mathrm{Tr}(\mathbf{P}_{AN}^H\mathbf{B}_e\mathbf{P}_{AN})}\times\frac{\mathrm{Tr}(\mathbf{P}_{AN}^H\mathbf{C}_e\mathbf{P}_{AN})}{\mathrm{Tr}(\mathbf{P}_{AN}^H\mathbf{C}_d\mathbf{P}_{AN})}\right)\nonumber\\
 &=\log_2\left(\frac{\mathbf{w}^H(\mathbf{I}_{N-1}\otimes \mathbf{B}_d)\mathbf{w}}{\mathbf{w}^H(\mathbf{I}_{N-1}\otimes\mathbf{B}_e)\mathbf{w}}\times\frac{\mathbf{w}^H(\mathbf{I}_{N-1}\otimes \mathbf{C}_e)\mathbf{w}}{\mathbf{w}^H(\mathbf{I}_{N-1}\otimes\mathbf{C}_d)\mathbf{w}}\right),
\end{align}
where $\mathbf{w}\triangleq \mathrm{vec}(\mathbf{P}_{AN})\in \mathbb{C}^{N(N-1)\times 1}$,
\begin{align}
\mathbf{B}_d=\frac{\beta_2^2}{\beta_1^2}\mathbf{H}_d+(\frac{\sigma^2}{\beta_1^2P_s}+\mathbf{h}^H(\theta_d)\mathbf{v}_d\mathbf{v}_d^H\mathbf{h}(\theta_d)  )\mathbf{I}_N,
\end{align}
\begin{align}
\mathbf{B}_e=\frac{\beta_2^2}{\beta_1^2}\mathbf{H}_e+(\frac{\sigma^2}{\beta_1^2P_s}+\mathbf{h}^H(\theta_e)\mathbf{v}_d\mathbf{v}_d^H\mathbf{h}(\theta_e)  )\mathbf{I}_N,
\end{align}
$\mathbf{C}_d=\mathbf{H}_d+\sigma^2(\beta_2^2P_s)^{-1}\mathbf{I}_N$, and $\mathbf{C}_e=\mathbf{H}_e+\sigma^2(\beta_2^2P_s)^{-1}\mathbf{I}_N$.

%are $N\times N$ matrices concerned with corresponding $\mathbf{v}_d$.

Consider that shrinking or stretching $\mathbf{P}_{AN}$ does not change the ratio value as shown in (\ref{Rs2}), the problem $\mathrm{(P1.1)}$ is equivalent to
\begin{align}\label{P1_2}
\mathrm{(P1.2):}&\max_{\mathbf{w}}~~~~\frac{\mathbf{w}^H(\mathbf{I}_{N-1}\otimes \mathbf{B}_d)\mathbf{w}}{\mathbf{w}^H(\mathbf{I}_{N-1}\otimes\mathbf{B}_e)\mathbf{w}}\times\frac{\mathbf{w}^H(\mathbf{I}_{N-1}\otimes \mathbf{C}_e)\mathbf{w}}{\mathbf{w}^H(\mathbf{I}_{N-1}\otimes\mathbf{C}_d)\mathbf{w}}.
\end{align}

Since $(\ref{P1_2})$ is a non-convex quadratic fractional function, $(\mathbf{I}_{N-1}\otimes \mathbf{B}_d)$, $(\mathbf{I}_{N-1}\otimes \mathbf{B}_e)$, $(\mathbf{I}_{N-1}\otimes \mathbf{C}_d)$ and $(\mathbf{I}_{N-1}\otimes \mathbf{C}_d)$ are $N(N-1)\times N(N-1)$ positive semi-definite matrices. $\mathbf{w}$ can be solved by utilizing GPI algorithm in \cite{Lee}. Then, $\mathbf{P}_{AN}$ is able to be reconstructed from $\mathbf{w}$.

%$\mathbf{Q}(\mathbf{w})=[\mathbf{w}^H(\mathbf{I}_{N-1}\otimes\mathbf{C}_d)\mathbf{w}](\mathbf{I}_{N-1}\otimes \mathbf{B}_e)+[\mathbf{w}^H(\mathbf{I}_{N-1}\otimes\mathbf{B}_e)\mathbf{w}](\mathbf{I}_{N-1}\otimes \mathbf{C}_d)$ and $\mathbf{V}(\mathbf{w})=[\mathbf{w}^H(\mathbf{I}_{N-1}\otimes\mathbf{C}_e)\mathbf{w}](\mathbf{I}_{N-1}\otimes \mathbf{B}_d)+[\mathbf{w}^H(\mathbf{I}_{N-1}\otimes\mathbf{B}_d)\mathbf{w}](\mathbf{I}_{N-1}\otimes \mathbf{C}_e)$.

\subsection{Optimize $\mathbf{v}_d$ for fixed $\mathbf{P}_{AN}$}
If the AN projection matrix $\mathbf{P}_{AN}$ in (\ref{P1}) is fixed , the optimization problem (\ref{P1}) is simplified to
\begin{align}\label{P1_3}
\mathrm{(P1.3):}&~~\max_{\mathbf{v}_d}~~~~~~~R_s(\mathbf{v}_d,~\text{fixed}~\mathbf{P}_{AN})\\ \nonumber
&~~\text{subject to}~~~~~\mathbf{v}_d^H\mathbf{v}_d=1
\end{align}
Observing (\ref{Rs1}), we find that above optimization problem is equivalent to
\begin{align}\label{P1_4}
\mathrm{(P1.4):}&~~\max_{\mathbf{v}_d}~~~~~~~~\frac{\mathbf{v}_d^H(\mathbf{H}_d+A_d\mathbf{I}_N)\mathbf{v}_d}{\mathbf{v}_d^H(\mathbf{H}_e+A_e\mathbf{I}_N)\mathbf{v}_d}\\ \nonumber
&~~\text{subject to}~~~~~~~~~~~\mathbf{v}_d^H\mathbf{v}_d=1
\end{align}

Actually, this is a generalized Rayleigh-Ritz ratio problem and the optimal $\mathbf{v}_d$ is the eigenvector corresponding to the largest eigenvalue of the matrix
\begin{equation}\label{v d}
(\mathbf{H}_e+A_e\mathbf{I}_N)^{-1}(\mathbf{H}_d+A_d\mathbf{I}_N).
\end{equation}

%\subsection{Convergence analysis}

\subsection{Initialization of $\mathbf{P}_{AN}$ and $\mathbf{v}_d$}
\begin{figure}[h]
 \centering
 \includegraphics[width=0.45\textwidth]{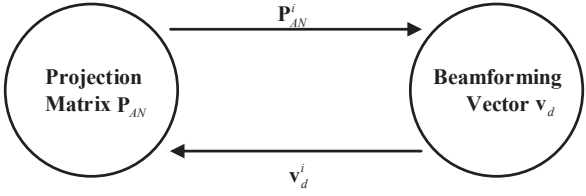}\\
 \caption{Schematic diagram of AIS}\label{Iteration}
\end{figure}

 Based on the two previous subsections, we propose an AIS by alternatively solving $\mathbf{P}_{AN}$ and $\mathbf{v}_d$ to further improve secrecy rate $R_s$ as shown in Fig.~\ref{Iteration}. The basic idea is to apply the two steps in Subsections A and B individually and  repeatedly until the SR converges. The detailed process is as follows. Firstly, we take the initial values of $\mathbf{P}_{AN}$ and $\mathbf{v}_d$ being the associated eigen-vectors of the largest egienvalues of (\ref{initial P_AN}) and (\ref{initial v d}), respectively. Then,  we  compute $\mathbf{P}_{AN}$ by utilizing GPI algorithm in \cite{Lee} for a given fixed  $\mathbf{v}_d$. Subsequently, $\mathbf{v}_d$ is chosen to be the associated eigen-vectors of the largest egienvalues of matrix(\ref{v d}) under the condition $\mathbf{P}_{AN}$ is fixed. We repeat the process until the termination condition is satisfied.
%$R_s$ according to the relative pairs of $\mathbf{P}_ {AN}^{i}$ and $\mathbf{v}_d^{i}$ as the iteration structure

Generally, the convergence rate of the iterative algorithm in Fig.~\ref{Iteration}  depends intimately on the the initial value. In what flows, we focus on the initialization of $\mathbf{P}_{AN}$ and $\mathbf{v}_d$ using the leakage concept in accordance with (\ref{Rx_signal yd}) and (\ref{Rx_signal ye}).

From the aspect of interfering the eavesdropper, the AN along direction $\theta_e$  should be viewed a useful component for eavesdropper, and the leakage of AN to the desired direction $\theta_d$ is regarded as interference. Hence, the AN-to-leakage-plus-noise ratio (ANLNR) corresponding to $\mathbf{P}_{AN}$ is defined as
\begin{align}\label{SLNR(P_AN)}
ANLNR(\mathbf{P}_{AN})&=\frac{\alpha^2\beta_2^2P_s\mathbf{h}^H(\theta_e)\mathbf{P}_{AN}\mathbf{P}_{AN}^H\mathbf{h}(\theta_e)}{\alpha^2\beta_2^2P_s\mathbf{h}^H(\theta_d)\mathbf{P}_{AN}\mathbf{P}_{AN}^H\mathbf{h}(\theta_d)+\sigma^2}\nonumber\\
&=\frac{\mathrm{Tr}[\mathbf{P}_{AN}^H\mathbf{H}_e\mathbf{P}_{AN}]}  {\mathrm{Tr}[\mathbf{P}_{AN}^H(\mathbf{H}_d+\frac{\sigma^2}{\beta_2^2P_s}\mathbf{I}_N)\mathbf{P}_{AN}]}.
\end{align}
When  maximizing the above cost function $ANLNR(\mathbf{P}_{AN})$,  the  optimization variable $\mathbf{P}_{AN}$ is a matrix. To solve this problem, we convert matrix $\mathbf{P}_{AN}$ into a column vector $\mathbf{w}$  using $\text{vec}$ operation, and the associated cost function becomes
\begin{align}\label{SLNR(P_AN))_1}
ANLNR(\mathbf{w})=\frac{\mathbf{w}^H(\mathbf{I}_{N-1}\otimes\mathbf{H}_e)\mathbf{w}}  {\mathbf{w}^H[\mathbf{I}_{N-1}\otimes(\mathbf{H}_d+\frac{\sigma^2}{\beta_2^2P_s}\mathbf{I}_N)]\mathbf{w}}.
\end{align}
Therefore, by maximizing the above objective function $ANLNR(\mathbf{w})$, we have  the optimum $\mathbf{w}$ being the eigenvector corresponding to the largest eigenvalues of matrix
\begin{equation}\label{initial P_AN}
\left[\mathbf{I}_{N-1}\otimes(\mathbf{H}_d+\sigma^2(\beta_2^2P_s)^{-1}\mathbf{I}_N)\right]^{-1}(\mathbf{I}_{N-1}\otimes\mathbf{H}_e).
\end{equation}
This completes the initialization of $\mathbf{P}_{AN}$.

Similarly, as the desired user hopes that the confidential message $x$ should be leaked to the eavesdropper along the eavesdropper direction $\theta_e$ as little as possible, we define the confidential signal-to-leakage-plus-noise ratio (CSLNR) corresponding to $\mathbf{v}_d$ as
\begin{align}\label{SLNR(v_d)}
CSLNR(\mathbf{v}_d)
&=\frac{\beta_1^2P_s\mathbf{v}_d^H\mathbf{h}(\theta_d)\mathbf{h}(\theta_d)^H\mathbf{v}_d}  {\beta_1^2P_s\mathbf{v}_d^H\mathbf{h}(\theta_e)\mathbf{h}(\theta_e)^H\mathbf{v}_d+{\sigma^2}}\nonumber\\
&=\frac{\mathbf{v}_d^H\mathbf{H}_d\mathbf{v}_d}  {\mathbf{v}_d^H(\mathbf{H}_e+\frac{\sigma^2}{\beta_1^2P_s}\mathbf{I}_N)\mathbf{v}_d}.
\end{align}

Maximizing (\ref{SLNR(v_d)}) yields the initial value of $\mathbf{v}_d$ being the eigenvector corresponding to the largest eigenvalues of matrix
\begin{equation}\label{initial v d}
(\mathbf{H}_e+\sigma^2(\beta_1^2P_s)^{-1}\mathbf{I}_N)^{-1}\mathbf{H}_d.
\end{equation}

\begin{figure}[ht]
 \centering
 \includegraphics[width=0.3\textwidth]{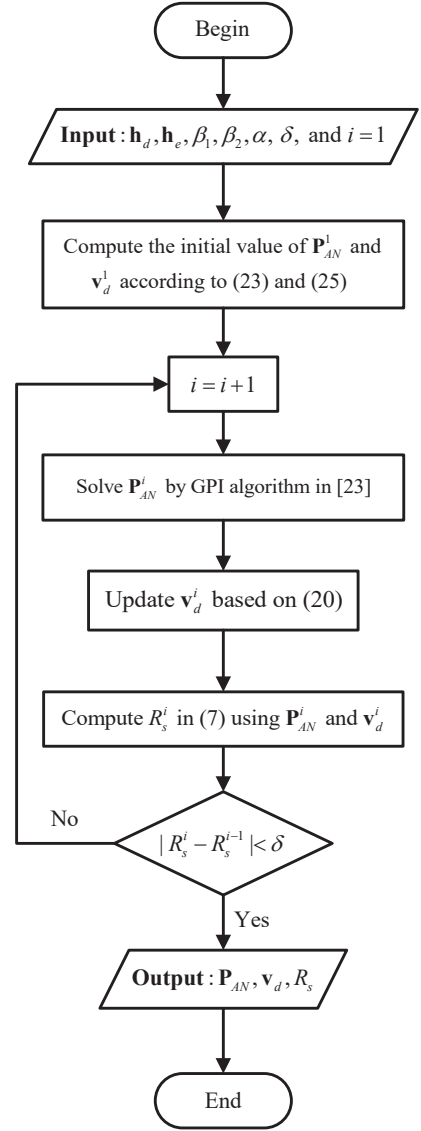}\\
 \caption{Flow graph of our proposed algorithm}\label{flow_chart}
\end{figure}

Finally,  the detailed implementation process of our proposed AIS scheme is summarized in Algorithm 1.

\begin{algorithm}[h]
\begin{algorithmic}
\STATE \textbf{Input:} $\mathbf{h}_d$, $\mathbf{h}_e$, $\beta_1$, $\beta_2$, $\alpha$ and $\delta$
\STATE \textbf{Output:} $\mathbf{P}_{AN}$, $\mathbf{v}_d$, $R_s$
\STATE \textbf{Initialization:} $i=1$, and compute the initial value of $\mathbf{P}_{AN}^1$ and $\mathbf{v}_d^1$ by using (\ref{initial P_AN}) and (\ref{initial v d}).
\REPEAT
\STATE 1. $i=i+1$.
\STATE 2. Update $\mathbf{P}_{AN}^i$ utilizing GPI algorithm in \cite{Lee};
\STATE 3. Update $\mathbf{v}_d^i$ based on (\ref{v d});
\STATE 4. Compute $R_s^i$ in (\ref{Rs1}) using updated $\mathbf{P}_{AN}^i$ and $\mathbf{v}_d^i$.
\UNTIL $|R_s^i-R_s^{i-1}|<\delta$.
\RETURN $\mathbf{P}_{AN}$, $\mathbf{v}_d$ and $R_s$.
\end{algorithmic}
\caption{Proposed iterative algorithm to solve $\mathbf{P}_{AN}$ and $\mathbf{v}_d$}\label{algorithm 1}
\end{algorithm}

In Algorithm 1, parameter $\delta$ is the tolerance factor. To make the above algorithm more clear, the corresponding detailed flow graph is also presented in Fig.~\ref{flow_chart}.

\begin{figure}[h]
 \centering
 \includegraphics[width=0.45\textwidth]{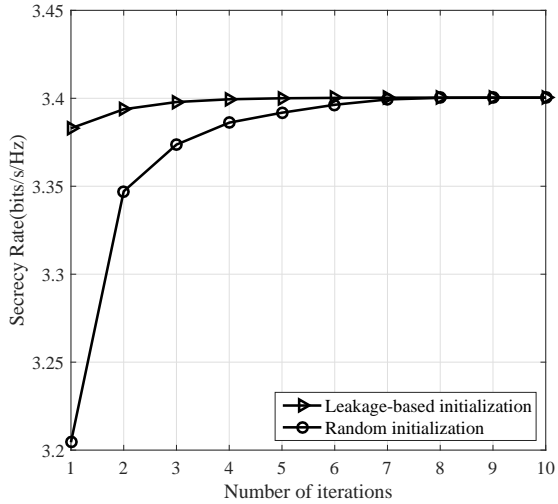}\\
 \caption{Curves of achievable secrecy rate versus the number of iterations (SNR=10dB)}\label{Ite_Num}
\end{figure}

\begin{figure}[h]
 \centering
 \includegraphics[width=0.45\textwidth]{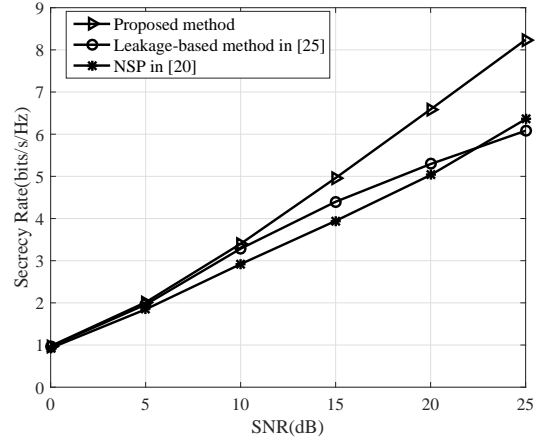}\\
 \caption{Achievable secrecy rate versus SNR of three different methods}\label{Sum_rate}
\end{figure}

\begin{figure}[h]
 \centering
 \includegraphics[width=0.45\textwidth]{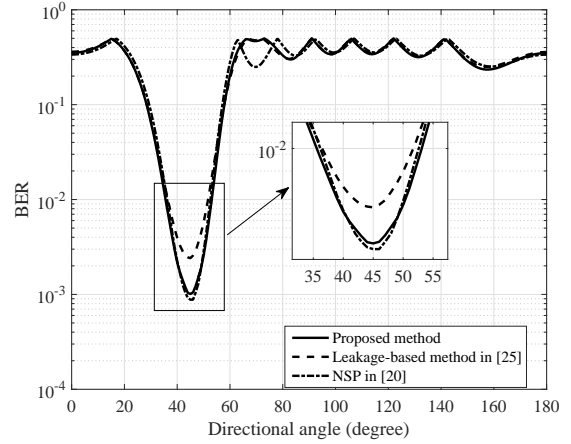}\\
 \caption{Curves of BER versus directional angles of three different methods}\label{Num_Iter}
\end{figure}

\section{Simulation,  Discussion, and Complexity Analysis}

%\subsection{Simulation Results and  Discussion,}
In this section, simulation results are presented to evaluate the performance of the proposed algorithm. The leakage-based method in \cite{Sadek} and NSP method in \cite{Dingy} are used For comparison, . The system parameters are chosen as:  $N=8$, PA factors $\beta_1=\sqrt{0.9}$ and $\beta_2=\sqrt{0.1}$,  $\theta_d=45^\circ$ and $\theta_e=70^\circ$, and QPSK.
\subsection{Simulation results and analysis}

Fig.~\ref{Ite_Num} shows the achievable SR versus the number of iterations between $\mathbf{P}_{AN}$ and $\mathbf{v}_d$  for  leakage-based and random initial values, respectively. Here, SNR is chosen to be 10dB. It is seen that the iterative algorithm with leakage-based solution as initialization value converges more rapidly than that with random initialization. As shown in Fig.~\ref{Ite_Num}, the former converges to a constant rate after 4 iterations, while the latter requires 7 iterations to converge. And the two initialization methods finally converge to the same SR limit value. This implies that the SLNR initialization can provide a faster convergence speed than the random initialization .

Fig.~\ref{Sum_rate} illustrates the achievable SR versus SNR of the proposed method, leakage-based method in \cite{Sadek} and NSP method in \cite{Dingy}. It can be seen that our proposed method performs better than the remaining two methods  in almost all SNR regions. With the increase of SNR, the secrecy rate gain over them achieved by the proposed method show a gradual growth trend. For example, at  SNR=15dB, the proposed method attains  an approximate ten-percent and twenty-percent rate improvements over leakage and NSP methods, respectively.

Fig.~\ref{Num_Iter} plots the bit error rate (BER)  versus SNR of the proposed method, leakage-based method and NSP. All the  three methods achieve their best BER performance along the desired direction $\theta_d=45^\circ$, and a sharp BER performance degradation appears once the desired receiver deviates from the main beam of the desired direction $45^\circ$. Both NSP and our proposed method have approximately the same  BER performance around the desired direction $45^\circ$. The main reason is that the proposed method reduces the effect of artificial noise on the desired direction by maximizing SR. The NSP  even makes AN vanish in the desired direction. However,  the conventional NSP one shows a better BER performance than the proposed method along the eavesdropper direction $\theta_e=70^\circ$ , which means  that the confidential messages  can be easily intercepted along the direction.
{\color{blue}{
\subsection{Complexity comparison and convergence analysis}

The complexities of the proposed Max-SR, NSP, and leakage-based methods are $O(I(N^6))$, $O(N^3)$, and $O(N^3)$ floating-point operations (FLOPs), respectively, where  $I$ stands for the number of iterations. Our proposed method is due to the fact that the optimization variable $\mathbf{w}\triangleq \mathrm{vec}(\mathbf{P}_{AN})$ in (13) is an  $N(N-1)$-D column vector.  Clearly, the proposed method has much higher complexity than NSP and leakage-based methods. Both NSP and leakage-based methods have the same order of complexity.

For the aspect of convergence, we propose two different choices of initial values in Section III: random and leakage-based. From Fig.~4,  the former requires 7 iterations to converge the limit and the latter needs only 4 iterations. Clearly, the leakage-based initialization approximately  doubles the convergence rate compared to the random initialization. This means a good initial choice accelerates the convergence of our proposed method.
}}
%\begin{table}[h]
%\centering
%\caption{Complexity comparison(FLOPs).}
%\newcommand{\tabincell}[2]{\begin{tabular}{@{}#1@{}}#2\end{tabular}}
%\begin{footnotesize}
%\begin{tabular}{|l|c|c}%{lccc}±íʾ¸÷ÁÐÔªËضÔÆ뷽ʽ£¬left-l,right-r,center-c
%\hline%±íʾÔÚ´ËÐÐÏÂÃæ»­Ò»ÌõºáÏß
%Methods & Complexity
%\\
%\hline
%Proposed Max-SR
%& $O(I(N^6)$ \\
%\hline
%Leakage
%& $O(N^3)$\\
%\hline
%NSP
%& $O(N^3)$\\
%\hline
%\end{tabular}
%\end{footnotesize}
%\end{table}

\section{Conclusion}
    In this paper, we have investigated three beamforming schemes including the proposed GPI-based Max-SR, leakage, and NSP. Compared to the last two methods {\color{blue}{based on NSP and leakage}}, the proposed GPI-based Max-SR method achieved a substantial SR improvement in the medium and large SNR regions. In particular, as SNR increases, the SR performance gain increases. Additionally, by an appropriate choice of initialization value, for example, with the leakage-based solution as initial value of GPI, the proposed method required only 4 iterations to converge to the limit value of SR. The random initialization requires 7 iterations. Thus, the leakage-based initialization saves about 40-percent computational amount over random one. {\color{blue}{In the coming future,  the proposed scheme will be potentially applied to the following diverse applications such as mmWave communications (massive MIMO), IoT, UAV, satellite communications, and flying ad-hoc networks.}}

% Note that IEEE does not put floats in the very first column - or typically
% anywhere on the first page for that matter. Also, in-text middle ("here")
% positioning is not used. Most IEEE journals use top floats exclusively.
% Note that, LaTeX2e, unlike IEEE journals, places footnotes above bottom
% floats. This can be corrected via the \fnbelowfloat command of the
% stfloats package.

\ifCLASSOPTIONcaptionsoff
  \newpage
\fi

\bibliographystyle{IEEEtran}
\bibliography{IEEEabrv,Security_Rate_bib}

\end{document}